\documentclass{article}
\usepackage{dcase2022,amsmath,graphicx,url,times,booktabs, tabularx}
\usepackage{color,makecell,bbm,amssymb,multirow,rotating,diagbox}
\usepackage{utfsym}

\title{Detect What You Want: Target Sound Detection}
\name{Helin Wang$^{1}$$^{\dagger}$\thanks{$^{\dagger}$ Indicates equal contribution.}, Dongchao Yang$^{1}$$^{\dagger}$, Yuexian Zou$^{1}$$^{*}$\thanks{$^{*}$ Corresponding Author: zouyx@pku.edu.cn}\thanks{This paper was partially supported by Shenzhen Science \& Technology Research Program (No: GXWD20201231165807007-20200814115301001; No: JSGG20191129105421211) and NSFC (No: 62176008).}, Fan Cui$^{2}$, Yujun Wang$^{2}$}
\address{$^1$ADSPLAB, School of ECE, Peking University, Shenzhen, China \\
         $^2$ Xiaomi Corporation, Beijing, China}
\begin{document}

\ninept
\maketitle

\begin{sloppy}

\begin{abstract}
Human beings can perceive a target sound type from a multi-source mixture signal by the selective auditory attention, 
however, such functionality was hardly ever explored in machine hearing.
This paper addresses the target sound detection (TSD) task, 
which aims to detect the target sound signal from a mixture audio when a target sound’s reference audio is given.
We present a novel target sound detection network (TSDNet) which consists of two main parts: A conditional network 
which aims at generating a sound-discriminative conditional embedding vector representing the target sound,
and a detection network which takes both the mixture audio and the conditional embedding vector as inputs and produces the detection result of the target sound.
These two networks can be jointly optimized with a multi-task learning approach to further improve the performance.
In addition, we study both strong-supervised and weakly-supervised strategies to train TSDNet and propose a data augmentation method by mixing two samples.
To facilitate this research, we build a target sound detection dataset (\textit{i.e.} URBAN-TSD) based on URBAN-SED and UrbanSound8K datasets, 
and experimental results indicate our method could get the segment-based F scores of 76.3$\%$ and 56.8$\%$ on the strongly-labelled and weakly-labelled data respectively.
\end{abstract}


\section{Introduction}
\label{sec:intro}
Human beings has the ability to focus auditory attention on a particular sound
in a multi-source environment,
however, there were few studies in this area for machine hearing.
In this paper, we initially define a target sound detection (TSD) task, 
which aims to recognize and localize target sound source within a mixture audio given a reference audio or/and a sound label. 
For example, the violin sound can be detected within a concert recording and the talking sound can be detected in a noisy cafe environment.
TSD has a lots of potential applications, 
such as noise monitoring for smart cities \cite{bello2018sonyc}, 
bioacoustic species and migration monitoring \cite{stowell2015acoustic} and large-scale multimedia indexing \cite{hershey2017cnn}.
To the best of our knowledge, this paper is the first attempt that explicitly tackles this problem.

There is one similar task with TSD, \textit{i.e.} sound event detection (SED).
SED aims to classify and localize all pre-defined sound events (\textit{e.g.}, train horn, car alarm) within an audio clip,
which has been widely studied \cite{dinkel2021towards,lin2020specialized,kong2020sound,kong2019weakly}. Compared to SED, TSD only
focuses on detecting the event that we care about and ignores other events. Furthermore, TSD does not require to pre-define categories set, so it can be easy to extend to open domain detection. Other related tasks about extracting the target signal are speaker extraction \cite{wang2018voicefilter,ge2021multi} 
which extracts the target speech from a mixture speech given a reference utterance of the target speaker, 
and acoustic events sound selection problems \cite{ochiai2020listen}.
Different from them, our work focuses on the detection task, 
which is more suitable for many multimedia retrieval applications and the training data is easier to obtain.

To solve the TSD task,
we propose a target sound detection network (\textit{i.e.} TSDNet),
and treat TSD as a binary classification problem for each frame of the audio,
where the positive class is the sound event of interest, 
and the negative class is formed by the combination of all foreground and background interfering events and noises. By using reference audio, 
TSDNet can focus on the target sound and ignore other interference. 
More specifically, 
TSDNet is composed of a conditional network which is used to generate a sound-discriminative conditional embedding vector from the reference audio, 
and a detection network which is applied to obtain binary-classification results at each frame with the conditional embedding vector and mixture audio as inputs.
In order to get more sound-discriminative conditional embedding vector for TSDNet, the conditional network is jointly optimized with both the TSD task and a sound event classification task. 
We further explore a data augmentation method for TSD, which randomly mixes two training samples to form a new training sample.
In addition, we explore to train the TSDNet on both strong- and weakly-supervised TSD tasks. 
Here, weakly-supervised TSD task means the dataset only provides the presence or absence of target sound within the mixture sound, 
but not any timestamp information, which is more challenging.

Our contributions can be summarized as follows:
(1) We propose a novel network for TSD that can be trained with both strong-labelled and weakly-labelled data.
(2) We explore the jointly training method to get the robust conditional embedding vector,
and propose a data augmentation method for the TSD task.
(3) We establish a dataset for TSD, and our method achieves the segment-based F measures of 76.3$\%$ and 56.8$\%$ on the strongly-labelled and weakly-labelled data respectively.

\section{Proposed Method}
The architecture of our proposed network (TSDNet) is shown in Figure \ref{fig:TEDNet}. 
TSDNet consists of two components:
a conditional network which inputs the reference audio and outputs a conditional embedding vector,
and a detection network
which uses the conditional embedding vector and the mixture audio to get the detection results. 
In this section, we will describe the details of the whole network.

\begin{figure}[t]
  \centering
  \includegraphics[width=\linewidth]{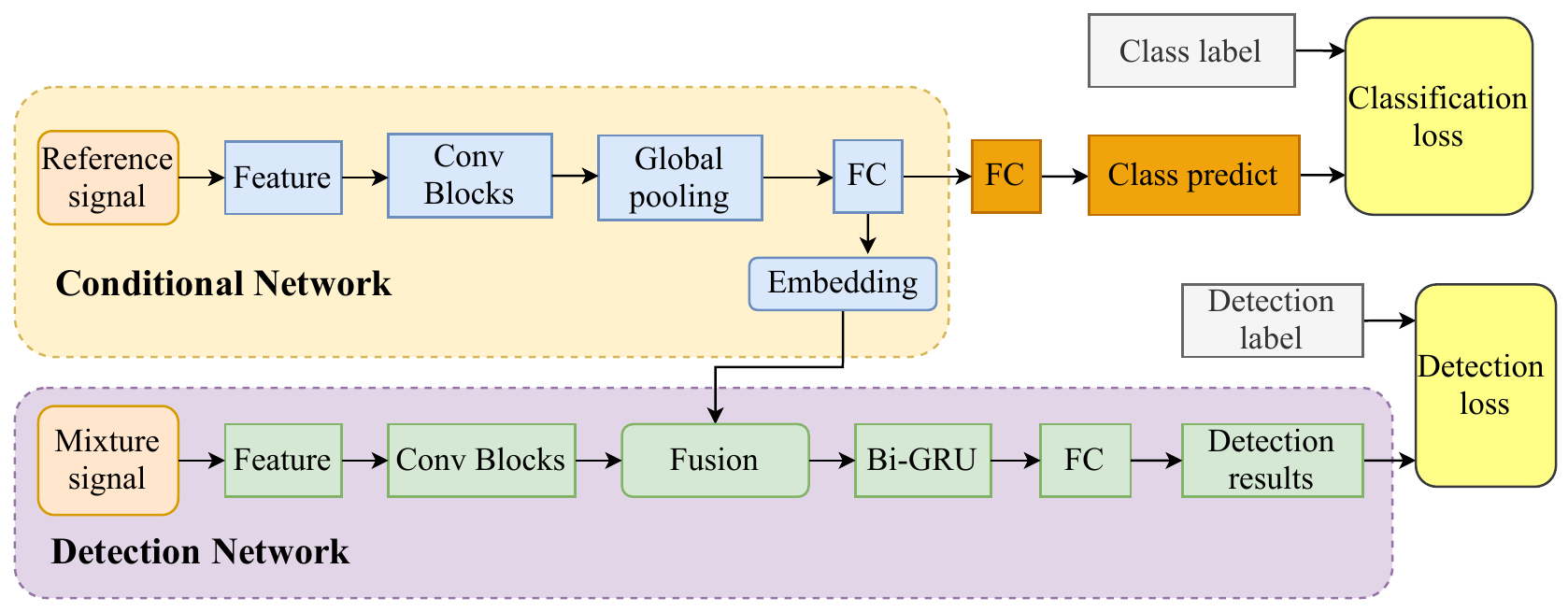}
  \caption{The architecture of our proposed TSDNet. Here, FC denotes the fully-connected layer.}
  \label{fig:TEDNet}
  \vspace*{-\baselineskip}
\end{figure}

\subsection{Conditional Network}
The purpose of the conditional network is to produce a global conditional embedding vector to represent the reference information. The input to the conditional network can be either a reference audio or a specific label, or both. In this paper, we focus on a reference audio as the input, for the reason that it is more challenging but easier to be transferred to new classes.
Inspired by the powerful ability of extracting time-frequency robust features from audio with convolutional neural networks (CNNs) \cite{kong2020panns, wang2019environmental,wang2019affects,wang2020acoustic}, we apply a VGG-like CNN network for the conditional network,
which uses the spectrogram feature as input and consists of 4 convolutional blocks with
64, 128, 256 and 512 output channels, respectively. 
Each convolutional block contains 2 convolutional layers with kernel size
of $3 \times 3$, followed by downsampling with average pooling size
of $2 \times 2$. Batch normalization \cite{ioffe2015batch} and ReLU function \cite{nair2010rectified} are
applied to all the convolutional layers. 
Global pooling layer \cite{kong2020panns} containing a global max-pooling function and a global average-pooling function is applied to get the global feature vector, which is then fed to a fully-connected layer to get the global conditional embedding vector with a fixed dimension of 128.

\begin{figure}[t]
  \centering
  \includegraphics[width=0.7\linewidth,height=0.5\linewidth]{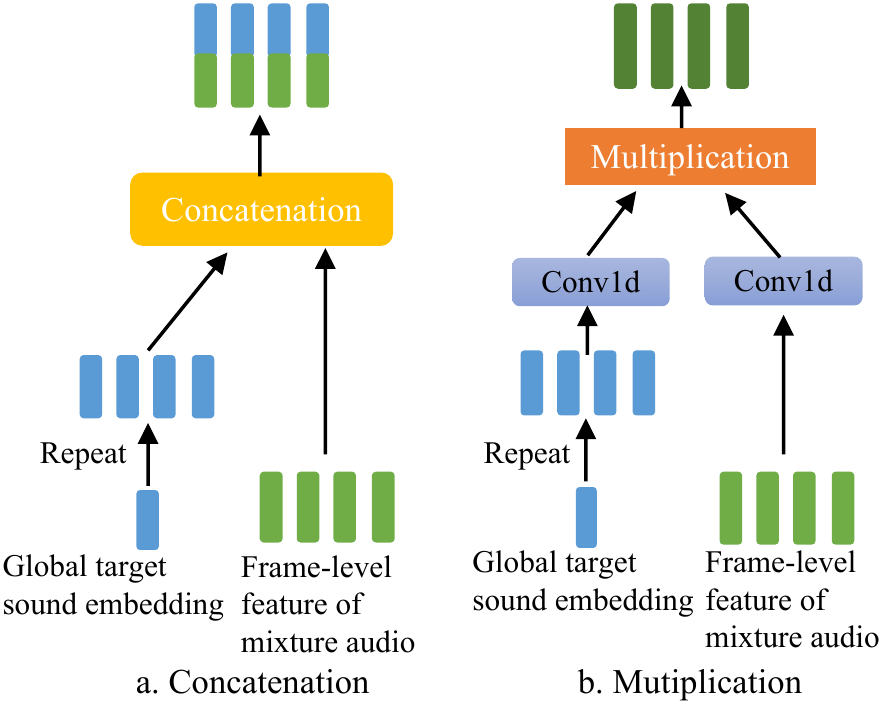}
  \caption{Two fusion modules.}
  \label{fig:fusion}
  \vspace*{-\baselineskip}
\end{figure}

\subsection{Detection network}
We have built two types of detection network: strong- and weakly-supervised network. The details are given as follows.\\
\textbf{Strong-supervised network} The strong-supervised detection network is based on the state-of-the-art work for weakly supervised sound event detection by Dinkel \textit{et al.} \cite{dinkel2021towards}. As shown in Figure \ref{fig:TEDNet}, the network has two inputs: the conditional embedding vector and the mixture audio, which is trained to minimize the difference between the frame-level prediction results and the ground-truth labels. To be more specific, the network is composed of 4 convolutional layers, 1 Bi-GRU layer, and 2 fully-connected layers, each with a LeakyReLU activation except the last layer, which has a sigmoid activation. 
Given the input feature of the mixture audio $\boldsymbol{x} \in \mathcal{R}^{t \times f}$, where $t$ and $f$ denote the number of frames and the dimension of each frame respectively, the network aims to predict frame-level probabilities $\hat{p}_i = \mathbb{P}(Y=k|X=x_i,\boldsymbol{e};\phi)$
where $\phi$ denotes the trainable parameters of detection network, $\boldsymbol{e}$ denotes the embedding from the conditional network and $x_i$ denotes the $i$-th frame of the mixture audio $\boldsymbol{x}$. 
Here, the value of $k$ is 0 or 1. Given the ground-truth label $p_i \in \{0,1\}$ for each frame, the strong-supervised network can be optimized by minimize the binary cross-entropy (BCE) loss function:
\begin{equation}\label{ssn}
\setlength{\abovedisplayskip}{4pt}
\setlength{\belowdisplayskip}{4pt}
\begin{split}
   \mathcal{L}_{sed} = \sum_{i=1}^{t}(-p_i\log\hat{p}_i-(1-p_i)\log(1-\hat{p}_i))
   \end{split}
\end{equation}
where $t$ indicates the number of frames. \\
\textbf{Weakly-supervised network} The difference between strong-supervised network and weakly-supervised network is that the latter needs a pooling layer to get the clip-level prediction. We add a linear softmax (LinSoft) pooling layer \cite{wang2019comparison} after the last layer of the strong-supervised detection network. It aims to predict a clip-level probability $\hat{P} = f_{LSP}(\hat{p}_1,\hat{p}_2, ..., \hat{p}_t)$ 
where $f_{LSP}(\cdot)$ denotes the linear softmax pooling function and  $T$ denotes the number of frames. 
Given the clip-level ground-truth label $P \in \{0,1\}$, BCE loss is also applied as the loss function:
\begin{equation}\label{wsn}
\setlength{\abovedisplayskip}{4pt}
\setlength{\belowdisplayskip}{4pt}
\begin{split}
   \mathcal{L}_{sed}^{'} = -P\log\hat{P}-(1-P)\log(1-\hat{P})
   \end{split}
\end{equation}

\subsection{Fusion Module}
As Figure \ref{fig:fusion} shows, we employ two fusion strategies to combine the conditional embedding and the feature of mixture audio. One is repeatedly concatenating the conditional embedding to the feature of mixture audio in each time frame. The other is projecting the conditional embedding and the feature of mixture audio to the same dimension by a 1-D convolutional layer, and then using multiplication operation to fuse them. 

\subsection{Mixup-TSD}
Following the advanced data augmentation methods \cite{zhang2017mixup,park2019specaugment,tokozume2018learning} for audio classification, 
we propose a data augmentation method based on the widely-used mixup \cite{zhang2017mixup} for TSD task.
The core idea is to create a new training sample by mixing a pair of two training samples.
More specifically,
we can generate a new training sample $(M_{new},R_{new},y_{new})$ from the data and label pair $(M_1,R_1,y_1)$ and $(M_2,R_2,y_2)$ by the following equation.
\begin{align}\label{mixup}
\setlength{\abovedisplayskip}{4pt}
\setlength{\belowdisplayskip}{4pt}
X_{new} &= \lambda X_1+(1-\lambda X_2)\\
R_{new} &= \lambda R_1+(1-\lambda R_2)\\
y_{new} &= \lambda y_1+(1-\lambda y_2)
\end{align}
where $M_1$ and $M_2$ are mixture audios, $R_1$ and $R_2$ are reference audios, $y_1$ and $y_2$ are the corresponding labels.
Such Mixup-TSD works by generating lots of new training samples, 
and particularly,
the target sound turns to be multi-label event instead of single-label event.
We argue that this method can perform well at the early training stage,
but needs a fine-tuning stage to fit for the single-label test.
In practice, we set a ratio $\alpha$ to control whether to use Mixup-TSD while training,
which means Mixup-TSD is applied with a probability of $\alpha$.
We set a linear decay for $\alpha$ from $0.3$ to $0$ during the whole training iterations.

\subsection{Jointly Training}
To get a more sound-discriminative conditional embedding vector for TSDNet,
we propose a jointly optimizing method for the conditional network via a multi-task manner. As show in Figure \ref{fig:TEDNet},
the feature vector after the global pooling layer is fed to a fully-connected layer and a softmax nonlinearity to get the classification results.
The conditional network can be optimized by minimizing the cross entropy loss between the predicted label and the ground truth label of the reference audio, along with the original detection task.
The whole loss function is defined by formula (\ref{Mul-task loss}). 
\begin{equation}\label{Mul-task loss}
\setlength{\abovedisplayskip}{4pt}
\setlength{\belowdisplayskip}{4pt}
\begin{split}
   \mathcal{L}_{\mathit{total}} = \mathcal{L}_{\mathit{sed}} + \mathcal{L}_{\mathit{cls}}
   \end{split}
\end{equation}
where $\mathcal{L}_{\mathit{sed}}$ denotes the loss of the detection task, and $\mathcal{L}_{\mathit{cls}}$ denotes the loss of the classification task, which is the cross-entropy loss between the ground-truth class labels and the predicted class labels. 

\begin{figure}[t]
  \centering
  \includegraphics[width=0.9\linewidth,height=0.45\linewidth]{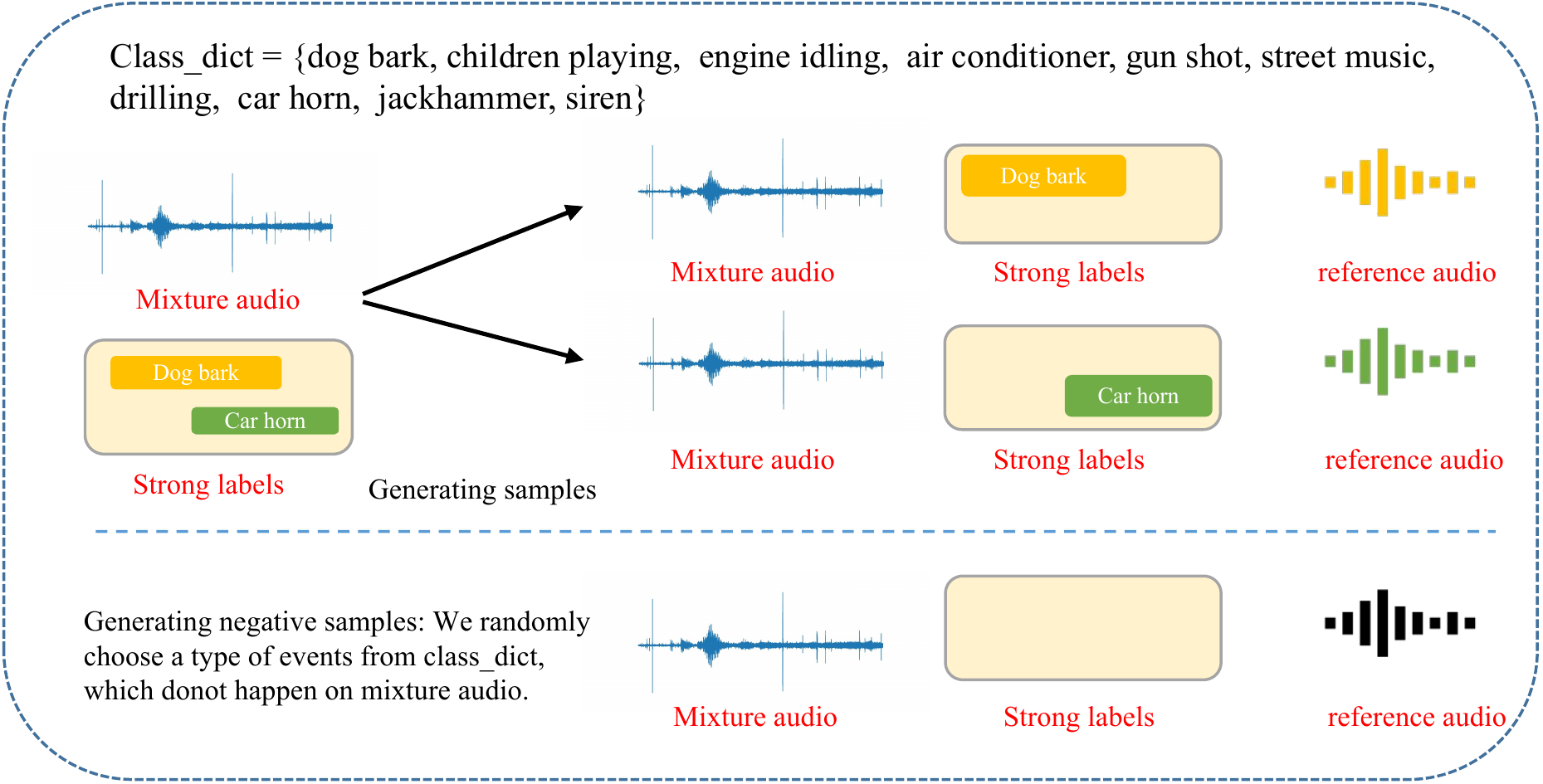}
  \caption{The process of data generation.}
  \label{fig:data}
  \vspace*{-\baselineskip}
\end{figure}

\section{URBAN-TSD Dataset}
As there was no dataset for TSD task,
we build a dataset called URBAN-TSD based on URBAN-SED \cite{salamon2017scaper} and UrbanSound8K datasets \cite{salamon2014dataset}. 
URBAN-SED is a sound event detection dataset with an urban setting, containing 10 event labels. 
This dataset’s source material is the UrbanSound8K dataset containing 27.8 hours of data split into about 4-second clip segments. 
The URBAN-SED dataset encompasses 10,000 soundscapes generated using the Scaper soundscape synthesis library \cite{salamon2017scaper}, 
which have been split into 6000 training, 2000 validation and 2000 test clips. 
The dataset contains mostly 10-second excerpts with strong labels, 
whereas each clip contains between one and nine events. 
We establish two types of dataset: strong-labelled dataset (URBAN-TSD-strong) and weakly-labelled dataset (URBAN-TSD-weak). 
The details of the number of samples are shown in Table~\ref{tab:data}.
\subsection{URBAN-TSD-strong}
We have built two strong datasets: URBAN-TSD-strong and URBAN-TSD-strong+. The details are given as follows.\\
\textbf{URBAN-TSD-strong dataset:} In this dataset, a sample includes three parts: a mixture audio, 
a reference audio and the strong label (the timestamp information of target sound). 
Mixture audios come from URBAN-SED dataset, and reference audio comes from UrbanSound8K. 
As shown in Figure~\ref{fig:data}, if there are $N$ sound events in the mixture audio, we can generate $N$ positive samples. 
For each positive sample, we randomly choose another sample that is in the same class from the UrbanSound8K as the reference audio.

\noindent
\textbf{URBAN-TSD-strong{+} dataset:} To further verify the ability of the model facing all negative frames, 
we generate samples that do not contain the target sound, 
which are called as negative samples. 
We add those samples because the mixture audio may not contain our target sound in the real world. 
The process of generating negative samples can summarize as:
For any mixture audio from URBAN-SED, 
we randomly choose a reference audio whose sound events do not happen in the mixture audio. 
The timestamp information (label) of negative samples are all set as 0. 
\subsection{URBAN-TSD-weak}
We also build a weakly-labelled dataset which is easier to obtain but more challenging. 
Comparing with strong-labelled dataset, 
there is no onset or offset time provided. 
The label is set as 1 (which indicates target sound happens in the mixture audio) 
or 0 (which indicates target sound does not happen in the mixture audio) for the whole audio clip. 
Similarly, for any mixture audio from URBAN-SED, if $N$ events happen, 
we can generate $N$ positive samples. 
For weakly-labelled dataset, negative samples are very important for training, 
and we generate the same number of negative samples as positive samples.

\begin{table}[t] \centering
\caption{The number of strong- and weakly-labelled data.}
\label{tab:data}
\begin{tabular}{cccc}
\hline
Type      & Strong & Strong{+} & Weak \\ \hline
Training     & 23106        & 29106                      & 41059        \\ \hline
Validation & 7681         & 9681                       & 13661        \\ \hline
Test      & 7702         & 9702                       & 13682        \\ \hline
\end{tabular}
\end{table}
\begin{table}[t] \centering
\caption{The performance comparison between TSDNet and other SED methods. Note that multiplication fusion is applied.}
\label{tab:my-table0}
\begin{tabular}{cc}
\hline
Method         & Segment-based F1  \\ \hline
CRNN \cite{martin2019sound} & 64.7                              \\
CDur \cite{dinkel2021towards}           & 64.8                               \\
CTrans \cite{miyazaki2020weakly}          & 64.51                   \\
SEDT-AQ-FT-P3 \cite{ye2021sound}          & 65.77                               \\
Ours           & \textbf{73.1}                             \\ 
\hline
\end{tabular}
\end{table}
\section{Experiments}
\subsection{Experimental setups}
\textbf{Training the conditional network:} We pre-train the conditional network with the classification task on the UrbanSound8K dataset. All the raw audios are down-sampled to 44.1kHz and applied a Short Time Fourier Transform (STFT) with a window size of 400 samples, followed by a Mel-scaled filter bank on perceptually weighted spectrogram. After that, we get log-mel spectrogram and MFCC feature, and concatenate them as the input, which is the state-of-the-art solution\footnote{https://www.kaggle.com/adinishad/urbansound-classification-with-pytorch-and-fun/notebook}. 
The Adam optimizer \cite{kingma2015adam} is used for a total of 50 epochs, with an initial learning rate of $1 \times 10^{-3}$.\\
\textbf{Training the detection network:} All the raw audios are down-sampled to 22.05kHz and applied a Short Time Fourier Transform (STFT) with a window size of 2048 samples, followed by a Mel-scaled filter bank on perceptually weighted spectrogram. This results in 64 Mel frequency bins and around 50 frames per second. The Adam optimizer \cite{kingma2015adam} is used for 100 epochs, with an initial learning rate of $1 \times 10^{-3}$. Note that we only update the detection network with the detection loss.\\
\textbf{Mixup-TSD:} For all the experiments, we use the Mixup-TSD method on the spectrogram level as we find spectrogram-level method works better than waveform-level method. 
Under otherwise stated, the linear decay is set for $\alpha$ from $0.3$ to $0$. \\
\textbf{Metrics:} We use the segment-based F-measure \cite{mesaros2016metrics} as the evaluation metric, which is the most commonly used metrics for sound event detection. All the F-scores are macro-averaged.
\begin{table}[t] \centering
\caption{The segment-based F-measure ($\%$) of TSDNet with different fusion strategies. We carry out the experiments for three
times and report the mean and standard deviation values.}
\label{tab:my-table-TEDNet-compare}

\begin{tabular}{|c|c|c|c|}
\hline
\textbf{Method}   & \textbf{strong}      & \textbf{strong+}     & \textbf{weak}        \\ \hline
Concatenation   & 73.1$\pm$0.61         & 67.2$\pm$0.41        & 52.8$\pm$0.53          \\ \hline
Multiplication & \textbf{73.1$\pm$0.19} & \textbf{69.3$\pm$0.23} & \textbf{53.0$\pm$0.32} \\ \hline
\end{tabular}
\end{table}
\begin{table}[t] \centering
\caption{The segment-based F-measure ($\%$) of different settings for Mixup-TSD. Here, multiplication fusion is applied.}
\label{tab:my-table}
\begin{tabular}{|c|c|c|c|}
\hline
\textbf{Mixup-TSD setting}   & \textbf{strong}      & \textbf{strong+}     & \textbf{weak}        \\ \hline
fixed ratio 0 & 73.1 & 69.3 & 53.0 \\ \hline
fixed ratio 0.2 & 74.9 & 71.1 & 55.2 \\ \hline
fixed ratio 0.5 & 74.2 & 70.8  & 54.7  \\ \hline
fixed ratio 0.8 & 74.0 & 70.4 & 53.6 \\ \hline
fixed ratio 1.0 & 73.8 & 70.2 & 53.2 \\ \hline
linear decay ratio & \textbf{75.1} & \textbf{71.6} & \textbf{55.8} \\ \hline
\end{tabular}
\end{table}

\subsection{The performance comparison between TSDNet and other SED methods}
In this part, we conduct experiments on URBAN-TSD-strong dataset to validate the effectiveness of our proposed method (TSDNet). We choose four previous state-of-the-art SED methods on URBAN-SED dataset. As for the baselines, we follow the model architecture in \cite{miyazaki2020weakly} to build the the Transformer-based model, which is referred to as CTrans in this paper. CDur \cite{dinkel2021towards} is the backbone of our TSDNet's detection network. Table 2 reports the results. We can see that our TSDNet significantly improve the performance due to introducing the reference audio. 
\subsection{Experimental results of different fusion strategies}
We evaluate our baseline method with different fusion strategies.
Table 3 reports the segment-based F-measure on the three datasets.
The multiplication gets larger mean values and smaller standard deviation values in all situations,
which is a more effective fusion method than the concatenation. In addition, TSDNet performs the worst on the URBAN-TSD-weak dataset,
for the reason that no timestamp information is available during training.

\subsection{Experimental results of Mixup-TSD}
We further evaluate the Mixup-TSD method on three datasets, 
and report the performance in Table 4.
If no augmentation is used,
TSDNet can obtain the segment-based F-measures of $73.1\%$, $69.3\%$ and $53.0\%$ respectively.
We can see that the Mixup-TSD method can significantly improve the performance over the baseline under all settings.
More specifically,
for a fixed ratio $\alpha$,
the performance tends to decrease with the ratio increasing,
and the best performance can be achieved when the ratio is $0.2$.
We argue that the Mixup-TSD method can perform well at the early training stage,
but it will change the detection target from single-label event to multi-label event so we need a fine-tuning stage to fit for the single-label test.
With a linear decay ratio from $0.3$ to $0$,
we can get the segment-based F-measures of $75.1\%$, $71.6\%$ and $55.8\%$ respectively.

\begin{table}[t] \centering
\caption{The performance of jointly training. 
Note that multiplication fusion and Mixup-TSD are applied.}
\label{tab:my-table-joint-train}
\begin{tabular}{cccc}
\hline
\textbf{Jointly training}   & \textbf{Strong}  & \textbf{Strong{+}} & \textbf{Weak}  \\ \hline
\usym{2613}  & 73.1  &69.3  & 53.0  \\ \hline
\checkmark & \textbf{76.3}  & \textbf{72.2}  & \textbf{56.8}   \\ \hline
\end{tabular}
\end{table}
\begin{table}[t] \centering
\caption{Experimental results of open domain. Note that multiplication fusion is applied.}
\label{tab:my-table-open}
\begin{tabular}{cccc}
\hline
\textbf{Jackhammer}  & \textbf{Siren}  & \textbf{Street\_music} & \textbf{Average}  \\ \hline
40.9 & 42.2  & 48.0  & 43.7  \\ \hline
\end{tabular}
\end{table}


\subsection{Experimental results of jointly training}
In addition, we explore whether jointly training can further improve the performance of target sound detection. Firstly, we initialize the TSDNet with a pre-trained model which is only trained by the detection task. Secondly, we fine-tune TSDNet by a multi-task manner introduced in Section 2.5. We set a learning rate of $1 \times 10^{-4}$ for the whole model, and the Adam optimizer \cite{kingma2015adam} is used for a total of 30 epochs. Experimental results are shown in Table \ref{tab:my-table-joint-train}.
We can see that TSDNet with jointly training obtains $76.3\%$, $72.2\%$ and $56.8\%$ respectively, which brings $4.3\%$, $4.2\%$ and $7.1\%$ improvement over the baseline respectively.
We find that the weak-supervised training still performs much worse than the strong-supervised training. We will study more effective weak-supervised methods in the future.
\subsection{Open domain target sound detection}
Table 6 shows the results of TSDNet evaluated on the open domain. We choose three events (jackhammer, siren, street\_music) as new classes, which does not occur in the training set. Specifically, we first exclude all of the data include the three events in the URBAN-TSD-strong dataset, named URBAN-TSD-strong-7. Then we train TSDNet on URBAN-TSD-strong-7. We can see that TSDNet can be applied to the new class, \textit{e.g.}, the F-score of street\_music achieved 48.0\% and the average F-score of the three new classes achieved 43.7\%. We believe that adding more training data can further improve the performance of new class. 
\section{Conclusions}
In this paper, we proposed a target sound detection network (TSDNet) which can be trained with both strong-supervised learning and weakly-supervised learning manners. In addition, our proposed Mixup-TSD data augmentation method and jointly training strategy further improve the performance of TSDNet.
TSDNet is more applicable to real scenarios because it does not require prior knowledge about the number of sound events and easy to extend to the new events.
In the future, we will explore more robust system.
The source code and datesets of this work are released.\footnote{https://github.com/yangdongchao/Target-sound-event-detection}

\bibliographystyle{IEEEtran}
\bibliography{refs}

%
%
%
%
%
%
%
%
%

\end{sloppy}
\end{document}